\documentclass[11pt]{article}
\usepackage{indentfirst}
\usepackage{comment}
\usepackage{appendix}
\usepackage{graphicx}
\usepackage{amsmath}
\usepackage{amsfonts}
\usepackage{amssymb}
\usepackage{setspace}
\usepackage{booktabs}
\usepackage{cancel}
\usepackage{float}
\usepackage{sectsty}
\usepackage{bm,bbm}
\usepackage[T1]{fontenc}
\usepackage{textcomp}
\usepackage{listings}
\usepackage{caption}
\usepackage{subcaption}
\usepackage{stmaryrd} 
\usepackage{times}
\usepackage{kotex}

\usepackage[backend=bibtex,sorting=none,citestyle=numeric-comp,]{biblatex}
\usepackage{titlesec}
\usepackage{xcolor}
\usepackage{ulem} 
\usepackage{soul} 

\usepackage{enumitem}
\usepackage[font=footnotesize,format=plain,labelfont=bf]{caption}
\usepackage{multirow}
\usepackage{kotex}
\usepackage{xcolor}
\usepackage{makecell}
\usepackage{amsmath,amssymb,amsfonts,mathtools}


\usepackage{url}
\usepackage[colorlinks]{hyperref} 
\hypersetup{ 
    colorlinks=true,       
    linkcolor=red,          
    citecolor=blue,        
    filecolor=magenta,      
    urlcolor=cyan           
}
\newcommand{\fref}[1]{Fig.~\ref{#1}}
\newcommand{\sref}[1]{Section~\ref{#1}}

\newcommand{\btau}{{\boldsymbol\tau}}

\newcommand{\jkref}[1]{\textcolor{red}{(jk-ref)}}
\newcommand{\tobecite}[1]{\textcolor{red}{(to be cite)}}

\begin{document}

\begin{center}
\textbf{\Large Graph-Based Learning of Free Surface Dynamics in Generalized Newtonian Fluids using Smoothed Particle Hydrodynamics }

\bigskip

Hyo-Jin Kim$^{1}$, Jaekwang Kim$^{2}$, Hyung-Jun Park$^{3, *}$\\
\bigskip
\small{
\textit{
$^1$Department of Korean Medical Science, College of Korean Medicine, \\Kyung Hee University, Seoul, Republic of Korea\\
$^2$Department of Mechanical and Design Engineering, Hongik University, Sejong, Republic of Korea\\
$^3$School of Mechanical and Aerospace Engineering / Center for Aerospace Engineering Research, \\Sunchon National University, Sunchon, Junnam, Republic of Korea \\
}

\bigskip

hjinkim@khu.ac.kr, jk12@hongik.ac.kr, hjpark89@scnu.ac.kr \\
$^*$ Corresponding author 
}
\end{center}

\bigskip

\begin{center}
\textbf{ABSTRACT }\\
\bigskip
\begin{minipage}{0.9\textwidth}
In this study, we propose a graph neural network (GNN) model for efficiently predicting the flow behavior of non-Newtonian fluids with free surface dynamics. The numerical analysis of non-Newtonian fluids presents significant challenges, as traditional algorithms designed for Newtonian fluids with constant viscosity often struggle to converge when applied to non-Newtonian cases, where rheological properties vary dynamically with flow conditions. Among these, power-law fluids exhibit viscosity that decreases exponentially as the shear rate increases, making numerical simulations particularly difficult. The complexity further escalates in free surface flow scenarios, where computational challenges intensify. In such cases, particle-based methods like smoothed particle hydrodynamics (SPH) provide advantages over traditional grid-based techniques, such as the finite element method (FEM).
Building on this approach, we introduce a novel GNN-based numerical model to enhance the computational efficiency of non-Newtonian power-law fluid flow simulations. Our model is trained on SPH simulation data, learning the effects of particle accelerations in the presence of SPH interactions based on the fluid’s power-law parameters. The GNN significantly accelerates computations while maintaining reliable accuracy in benchmark tests, including dam-break and droplet impact simulations. The results underscore the potential of GNN-based simulation frameworks for efficiently modeling non-Newtonian fluid behavior, paving the way for future advancements in data-driven fluid simulations.

\end{minipage}
\end{center}

\bigskip
\begin{minipage}{0.9\textwidth}
\end{minipage}

\newpage
\section{Introduction}

Many fluids encountered in fields such as chemical engineering, biofluid mechanics, and industrial processing are non-Newtonian, exhibiting diverse rheological characteristics~\cite{peng2014application}.
In particular, a significant number of these non-Newtonian flows involve a free surface—a feature common in industrial applications such as coating, inkjet printing, and additive manufacturing, where both complex rheology and interface dynamics play crucial roles~\cite{JKim:2019,JKim:2020,JKim:2021}.
For these types of flows, numerical approaches have received sustained attention, as analytical solutions are often intractable and experimental observations can be costly.
However, achieving accurate simulations of non-Newtonian fluids at reasonable computational costs remains a major challenge. This difficulty arises from the dynamic nature of their rheological properties that vary with local flow conditions.
Conventional computational fluid dynamics (CFD) algorithms, i.e., Navier-Stokes solvers or its variation, have been developed originally for Newtonian fluids with constant viscosity, they struggle to maintain numerical stability when applied to non-Newtonian cases~\cite{fortin1989new}. 

In response to these challenges, extensive research has focused on developing numerical methods tailored for non-Newtonian fluid flows, particularly those that capture complex rheological behaviors like shear-thinning and viscoplasticity.
The finite element method (FEM) has been widely adopted due to its flexibility in handling irregular geometries and complex boundary conditions ~\cite{Bathe1996, dvorkin1988non, wackerfuss2009mixed, kim2023evaluation, kim2023laser}.
Various stabilized FEM formulations have been introduced to model generalized Newtonian and viscoelastic fluids, though they often incur high computational costs~\cite{szadyNewMixedFinite1995,grilletModelingViscoelasticLid1999,varchanisNewFiniteElement2019,kimAdjointbasedSensitivityAnalysis2023a}. 
Similarly, the finite volume method (FVM), commonly used in commercial CFD solvers, has been extended to support power-law and Bingham-type models for simulating non-Newtonian flows in complex domains~\cite{neofytou3rdOrderUpwind2005,deCoupledFiniteVolume2016,deViscoelasticFlowSimulations2017,deComplexFluidsViscoelastic2022}. 
Collectively, these approaches are categorized as mesh-based methods, and they are often equipped with augmented numerical techniques, such as regularization, to stabilize the scheme and ensure the convergence of numerical solutions, even at the cost of computational efficiency or accuracy~\cite{JKim:2019}.

When the problem domain includes a free surface, obtaining a convergent solution using a mesh-based approaches becomes even more demanding.
In such scenarios, mesh-based algorithms also need to simultaneously handle complex material behavior and dynamically evolving interfaces, resulting in considerable numerical difficulty~\cite{zhang2024state}.
On the other hand, particle-based methods like smoothed particle hydrodynamics (SPH) have gained long attention for free surface problems, as they inherently handle complex interface dynamics without requiring specialized surface-tracking algorithms~\cite{parkNewSPHFEMCoupling2024,kimDirectImpositionWall2023,parkSegmentbasedWallTreatment2024}. 
This has motivated the non-Newtonian fluid mechanics community to embed non-Newtonian constitutive rheological models into SPH formulations to simulate shear-thinning, shear-thickening, and yield-stress fluids~\cite{shaoIncompressibleSPHMethod2003a,hosseiniFullyExplicitThreestep2007,xuSPHSimulationsThreedimensional2013,xuSPHSimulationsNonisothermal2025, Son:2023}. 
Moreover, parallelization techniques utilizing graphics processing units (GPUs) have also been explored to mitigate the associated computational costs~\cite{renImprovedParallelSPH2016}. However, these hardware-dependent approaches are limited by system-specific constraints and scalability issues.

With recent advances in machine learning and artificial neural networks for mechanics, data-driven approach have emerged as viable alternatives to overcome such issues~\cite{kimApplicationArtificialNeural2023a,kimDeepFluidsGenerative2019,caiPhysicsinformedNeuralNetworks2021, cho2024fault}.
In particular, graph neural networks (GNNs) have demonstrated strong effectiveness in learning and predicting the outcomes of particle-based fluid simulations.
Their graph-based structure closely resembles the local interactions among particles, making them well-suited for learning the result of particle-based method. For instance, previous studies have successfully applied GNNs to simulate Newtonian and multiphase flows, demonstrating notable generalization capabilities and substantial reductions in computational cost after training~\cite{Sanchez-Gonzalez2020,liGraphNeuralNetworkaccelerated2022a,zhangHybridMethodCombining2024}. 
To the author's knowledge, however, existing GNN-based fluid simulation models have only addressed the constant viscosity of Newtonian fluids

In this study, we propose a GNN-based simulation model for simulating flows of non-Newtonian power-law fluids. 
Among the various types of non-Newtonian rheological models, power-law fluids fall within the class of generalized Newtonian fluids, which are characterized as purely viscous fluids whose viscosity changes instantaneously with shear rate.
Depending on its model parameter, power-law fluids exhibit both shear-thinning and shear-thickening behaviors.  
Then, we augment the node features of the GNN with model parameters specific to power-law fluids and train the network using data generated from SPH simulations. The GNN learns to predict particle interactions without explicitly solving the momentum equations at each time step. A key advantage of this GNN-based approach is its greater flexibility in choosing the time-step size, which is typically constrained by the maximum particle viscosity in SPH implementations of power-law fluids. This flexibility leads to significant computational gains.

The remainder of this paper is organized as follows. \sref{sec:numerical_method} summarizes the physical formulation of power-law fluids and the SPH framework for variable viscosity 
The results of SPH simulations of power-law fluids flows will be used to train GNN models for power-law fluids. 
\sref{sec:GNN_model} details our GNN-based simulation model, covering the data generation process, model architecture, and training strategy. \sref{sec:result} presents the results and discussion, evaluating the model’s generalization capability and performance under various viscosity conditions through benchmark tests such as droplet impact and dam-break scenarios. Finally, \sref{sec:conclusions} summarizes the findings and outlines potential directions for future research.

\section{Power-Law Fluid Formulation and SPH Scheme} \label{sec:numerical_method}

This section outlines the physical formulation of power-law fluid flows and an extended SPH scheme that incorporates variable viscosity. The resulting SPH simulations serve as training data for the GNN-based model introduced in the next section.

\begin{figure}
\begin{center}
\includegraphics[width=0.65\textwidth]{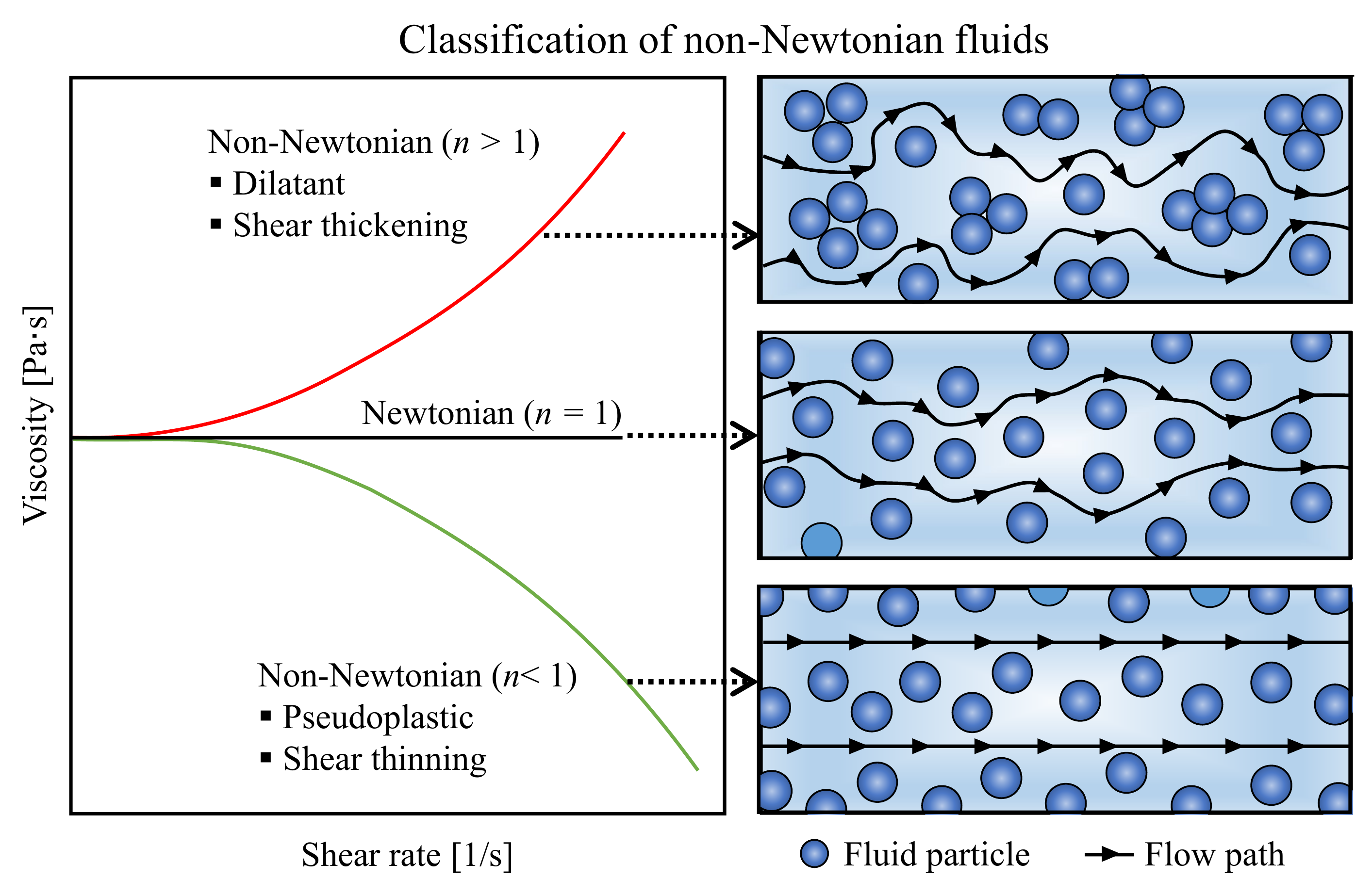}
\end{center}
\caption{Schematic illustration of shear-dependent viscosity behavior in power-law fluids as a function of shear rate. The flow behavior index $n$ governs the fluid response: $n$ < 1 corresponds to shear-thinning (pseudoplastic) fluids, 
$n$ = 1 represents Newtonian fluids with constant viscosity, and $n$ > 1 indicates shear-thickening (dilatant) fluids.}
\label{fig:nonNewtonian_schematic}
\end{figure}

\subsection{Governing equation for non-Newtonian fluid flows}

In this paper, we consider isothermal fluid flows
of a power-law fluid, which are governed by the the conservation of the mass 
\begin{equation}
    \frac{{D\rho }}{{Dt}} =  - \rho \nabla  \cdot {\bf{v}},
    \label{eqn:continuityEqn}
\end{equation}
and momentum
\begin{equation}
    \frac{D\mathbf{v}}{Dt}=-\frac{1}{\rho }\nabla p+\frac{1}{\rho }\nabla \cdot \boldsymbol {\tau }+{{\mathbf{F}}^{B}},
    \label{eqn:momentumEqn}
\end{equation}
where $\rho $ is fluid density, ${\bf{v}}$ is velocity, $p$ is pressure, $\boldsymbol{\tau}$ is the material stress tensor, and ${{\bf{F}}^B}$ is buoyancy force.

If the fluid is a Generalized 
Newtonian Fluid (GNF), the stress tensor $\boldsymbol{\tau}( \dot{\boldsymbol{\epsilon }})$ 
consists solely of viscous stress and is a nonlinear instantaneous function of the local strain rate 
\begin{equation*}
     \dot{\boldsymbol{\epsilon }}=\frac{1}{2}\left[ \nabla \mathbf{v}+\nabla {{\mathbf{v}}^{\text{T}}} \right].
\end{equation*}
Note that since the GNF model accounts only for the current strain rate, it cannot capture time-dependent behaviors such as fluid memory, relaxation, or viscosity changes over time.
More specifically, $\btau$ is given as 
\begin{equation}
    \btau = 2\mu({\dot{\gamma }})  \dot {\boldsymbol{\epsilon}},
    \label{eqn:GNFstress}
\end{equation}
where $\dot{\gamma}=\sqrt{2\dot{\boldsymbol{\epsilon }}:\dot{\boldsymbol{\epsilon }}} \;  (>0)$ is shear rate and $\mu (\dot\gamma)$ is the dynamics viscosity. 
Different forms of $\mu$ correspond to different types of GNF models.

In this work, we consider a power-law fluid 
which states $\mu$ as 
\begin{equation}
    \mu (\dot\gamma)=K{{{\dot{\gamma }} }^{n-1}},
    \label{eqn:powerLaw}
\end{equation}
where $K$ and $n$ are the model parameters of the power-law fluid which are called flow consistency index and the flow behavior index respectively. 
Note that $n$ indicates the degree to which the fluid deviates from Newtonian behavior.
For example, when $n=1$, the fluid exhibits Newtonian flow; when $n<1$, it shows shear-thinning behavior; and when $n>1$, it displays shear-thickening behavior. The conceptual diagram of flow characteristics based on the value of $n$ is shown in \fref{fig:nonNewtonian_schematic}.
In this paper, we consider a $K$ range of 0.1 to 10 and an $n$ range of 0.5 to 2.0.

Among the various GNF models defined by different forms of $\mu$, the power-law model is one of the simplest. Its two-parameter formulation allows for efficient training of GNN models with relatively small datasets.

\subsection{SPH for Generalized Newtonian flows}

\begin{figure}
\begin{center}
\includegraphics[width=0.75\textwidth]{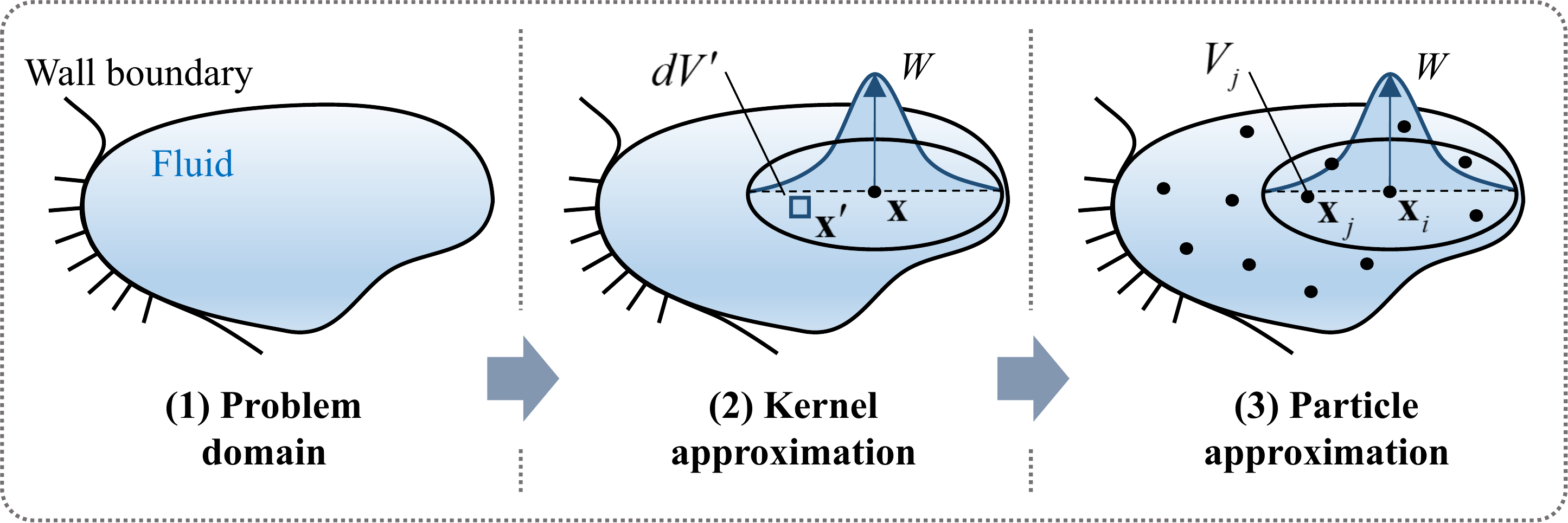}
\end{center}
\caption{Illustration of the SPH discretization process in three stages: (1) the continuous problem domain is defined; (2) kernel approximation is applied to express field quantities as weighted integrals; and (3) the integrals are discretized into particle summations over neighboring particles.
}\label{fig:schematic_illustration_SPH_discretization}
\end{figure}

SPH is a particle-based Lagrangian approach that solves the governing equations by calculating inter-particle interactions, updating particle states such as positions and velocities accordingly.
SPH performs the discretization process through two steps, as shown in \fref{fig:schematic_illustration_SPH_discretization}: (1) kernel approximation and (2) particle approximation. In the kernel approximation step, the physical information $f(\mathbf{x})$ at an arbitrary point $\mathbf{x}$ is determined by collecting information from surrounding points $\mathbf{x'}$ within a certain radius. This region, defined by the radius, is called the support domain. A function $W$ based on the distance between points $\mathbf{x}$ and $\mathbf{x'}$ is multiplied by the physical information $f(\mathbf{x'})$ at point $\mathbf{x'}$, and by integrating it within the support domain, $f(\mathbf{x})$ is interpolated. This process is expressed as follows:

\begin{equation}
    f(\mathbf{x})=\int_{\Omega }{f(\mathbf{{x}'})W(\mathbf{x}-\mathbf{{x}'},h)d{V}'},
    \label{eqn:kernelApproximation}
\end{equation}
where $f(\mathbf{x})$ and $f(\mathbf{x'})$ represent the physical values at positions $\mathbf{x}$ and $\mathbf{x'}$, respectively, $W$ is the kernel function, and $h$ is the smoothing length used to define the radius of the kernel function, ${\kappa}h$, where ${\kappa}$ is a scaling factor. In this paper, ${\kappa}$ is set to 2.

In the second step, the particle approximation, the integral expression in Eq. \eqref{eqn:kernelApproximation} is discretized using a finite number of particles and is replaced with a summation form. Here, the infinitesimal differential element $dV'$ becomes the volume of each particle, and the equation is expressed as follows:

\begin{equation}
    f({{\mathbf{x}}_{i}})=\sum\limits_{j}{f({{\mathbf{x}}_{j}})W({{\mathbf{x}}_{i}}-{{\mathbf{x}}_{j}},h)}\frac{{{m}_{j}}}{{{\rho }_{j}}},
    \label{eqn:particleApproximation}
\end{equation}
in which subscripts $i$ and $j$ denote the particle indices, where particle $i$ is the target particle and particle $j$ represents the particles located within the support domain of particle $i$. Note that the infinitesimal differential element $dV'$ is transformed into the volume of a particle, $\frac{{{m}_{j}}}{{{\rho }_{j}}}$, where $m_{j}$ is the mass of the particle $j$.

By applying Eq. \eqref{eqn:particleApproximation} to Eqs. \eqref{eqn:continuityEqn} and \eqref{eqn:momentumEqn}, the following discretized SPH equations are obtained:

\begin{equation}
    \frac{D{{\rho }_{i}}}{Dt}={{\rho }_{i}}\sum\limits_{j}{({{\mathbf{v}}_{i}}-{{\mathbf{v}}_{j}})\cdot {{\nabla }_{i}}W({{\mathbf{x}}_{i}}-{{\mathbf{x}}_{j}},h)\frac{{{m}_{j}}}{{{\rho }_{j}}}}+{{D}_{i}},
    \label{eqn:discretedConstitutiveEqn}
\end{equation}
\begin{equation}
    \frac{D{{\mathbf{v}}_{i}}}{Dt}=-\sum\limits_{j}{\left( \frac{{{p}_{i}}+{{p}_{j}}}{{{\rho }_{i}}{{\rho }_{j}}} \right){{\nabla }_{i}}W({{\mathbf{x}}_{i}}-{{\mathbf{x}}_{j}},h){{m}_{j}}}+\left\langle \frac{1}{\rho }\nabla \cdot \boldsymbol{\tau}  \right\rangle +\mathbf{F}_{i}^{B},
    \label{eqn:discretedmomentumEqn}
\end{equation}
where ${{D}_{i}}$ is the density diffusion term used to alleviate instability of the density field and $\left\langle \cdot  \right\rangle$ denotes the SPH discretization. The second term on the right-hand side of Eq. \eqref{eqn:discretedmomentumEqn} represents the material stress term, whose discretization for SPH will be explained later in this section.

Various SPH models exist depending on the approach used to calculate the pressure in Eq. \eqref{eqn:discretedmomentumEqn}. In this work, we employ the Weakly Compressible SPH model (WCSPH), which is the simplest and most widely used model. In WCSPH, pressure is computed in an extrapolated manner using the Tait equation, a form of the equation of state, as follows:
\begin{equation}
    p=\frac{{{\rho }_{0}}{{c}^{2}}}{\gamma }\left[ {{\left( \frac{\rho }{{{\rho }_{0}}} \right)}^{\gamma }}-1 \right],
\end{equation}
where ${\rho }_{0}$ is the reference density, $c$ is the speed of sound, and $\gamma$ is the adiabatic index, which is set to 7 in this paper. The speed of sound, $c$, is determined to maintain the incompressibility of the fluid.
Since pressure in WCSPH is calculated on the basis of the fluid's density, the density field becomes unstable in regions where fluid velocity changes rapidly, which in turn leads to an unstable pressure field. To mitigate this, the density diffusion term is widely used~\cite{marrone2011delta}, and in this paper, we adopt the $\delta$-SPH scheme proposed by Sun et al.~\cite{sun2017deltaplus}. In the $\delta$-SPH scheme, the density diffusion term ${{D}_{i}}$ is defined as follows:

\begin{gather}
    {{D}_{i}}=-2\delta hc\sum\limits_{j}{{{R}_{ij}}\frac{({{\mathbf{x}}_{i}}-{{\mathbf{x}}_{j}})\cdot {{\nabla }_{i}}W({{\mathbf{x}}_{i}}-{{\mathbf{x}}_{j}},h)}{{{({{\mathbf{x}}_{i}}-{{\mathbf{x}}_{j}})}^{2}}+{{(0.1h)}^{2}}}}\frac{{{m}_{j}}}{{{\rho }_{j}}} \\    
    \text{with} \quad
    {{R}_{ij}}=({{\rho }_{j}}-{{\rho }_{i}})-\frac{1}{2}\left( \left\langle \nabla \rho  \right\rangle _{j}^{L}+\left\langle \nabla \rho  \right\rangle _{i}^{L} \right)\cdot ({{\mathbf{x}}_{j}}-{{\mathbf{x}}_{i}}),
    \label{eqn:deltaScheme}
\end{gather}
in which $\delta$ is a parameter to control the strength of the density diffusion and $\left\langle \nabla \rho  \right\rangle ^{L}$ represents the re-normalized density gradient term~\cite{sun2017deltaplus}.

Lastly, to implement power-law fluid flows in SPH, an appropriate formulation of the material stress tensor term in the discretized momentum conservation equation (Eq. \eqref{eqn:discretedmomentumEqn}) is required. We express this term as follows~\cite{fourtakas2016modelling}: 

\begin{equation}
    \left\langle \frac{1}{\rho }\nabla \cdot \boldsymbol{\tau } \right\rangle =\sum\limits_{j}{\left( \frac{{{\boldsymbol{\tau }}_{i}}+{{\boldsymbol{\tau }}_{j}}}{{{\rho }_{i}}{{\rho }_{j}}} \right)\cdot }{{\nabla }_{i}}W({{\mathbf{x}}_{i}}-{{\mathbf{x}}_{j}},h){{m}_{j}}.
\end{equation}    

To utilize the constitutive equation of the power-law fluids, Eqs. \eqref{eqn:GNFstress} and \eqref{eqn:powerLaw}, we consider the discretized
form of the strain rate tensor as follows~\cite{kim2025comparative}

\begin{gather}
    \left\langle {\dot{\boldsymbol{\epsilon }}} \right\rangle =\frac{1}{2}\left[ \left\langle \nabla \mathbf{v} \right\rangle +{{\left\langle \nabla \mathbf{v} \right\rangle }^{\text{T}}} \right]\\    
    \text{with} \quad \left\langle \nabla \mathbf{v} \right\rangle =\sum\limits_{j}{\left( {{\mathbf{v}}_{j}}-{{\mathbf{v}}_{i}} \right)\otimes {{\nabla }_{i}}W({{\mathbf{x}}_{i}}-{{\mathbf{x}}_{j}},h)\frac{{{m}_{j}}}{{{\rho }_{j}}}}. 
\end{gather}
Once the strain rate is calculated, the shear rate, dynamic viscosity, and material stress can be determined sequentially.

\section{A GNN-based Predictive Model for power-law fluids}
\label{sec:GNN_model}

\begin{figure}
\begin{center}
\includegraphics[width=0.65\textwidth]{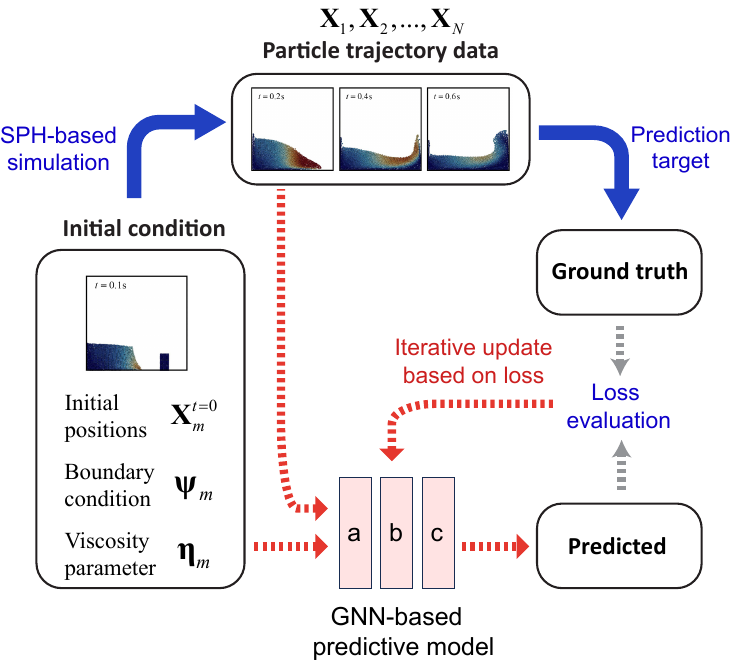}
\end{center}
\caption{Overall structure of the proposed framework.
}\label{fig:Model_overview}
\end{figure}

\subsection{Overview of the framework}

We begin with a broad overview of power-law fluid flow modeling with graph neural networks.
In our framework, transient flows are represented by the time evolution of fluid particle positions, with updates governed by velocities and accelerations arising from interactions with nearby particles.
Recognizing the graph-like structure—particles as nodes and interactions as edges, we design a GNN to learn these dynamics. 
The role of the GNN model is to predict the accelerations of individual particles. Training data are generated using the SPH method across various power-law fluid scenarios. Once trained, the model can generalize to predict particle dynamics under unseen physical conditions.
Our overall framework is illustrated in \fref{fig:Model_overview}. 
In the following subsections, we describe the structure of the SPH raw simulation data, the architecture of the graph neural network, the construction of the message-passing unit, and the training process.
Hyper parameters are introduced throughout this section, with the specific values used in this study summarized at the end.

\subsection{Particle data structuring and feature embedding}
\label{subsec:embedding}

Here, we describe the SPH particle data structure, which is used to train the GNN model, along with the data processing steps that transform the raw simulation outputs into embedded feature representations.
In our framework, the inputs of SPH simulations for flow scenario-$m$ are identified as (a) the two power-law model parameters $\bm{\eta}_{m}=\{K_m, n_m\}$, (b) the boundary particle labels $\bm{\Psi}_{m}  \in \mathbb{Z}^{N_p^m \times 1}$ and (c) initial SPH particle positions $\mathbf{X}_{m}^{t=0}$ at time $t=0$ of the total $N^p_m$ SPH particles. 

First, the boundary particle labeling $\bm{\Psi}_{m}$, which we write as 
\begin{equation}
\bm{\Psi}_{m}= \left[ \psi_{m(1)}, \psi_{m(2)}, \dots, \psi_{m(N_p^m)} \right],
\end{equation}
where $\psi_{m(i)}$ represents the label of the $i$-th particle,
characterizes the boundary conditions of the given flow case $m$. 
In particular, there are two types of particles: fluid particles, labeled as 0, which simulate fluid motion; and rigid particles, labeled as 1, which represent boundary walls or internal rigid obstacles within the simulation domain.
This categorical data $\bm{\Psi}_{m}$ is embedded into a higher-dimensional feature space, of dimension $\mathsf{N}_e$, 
an embedding transformation $\phi^b$ is applied:
\begin{equation}
\phi^b: \mathbb{Z}^{N^p_m \times 1} \to \mathbb{R}^{N_m^p \times \mathsf{N}_e}
\label{eqn:embeddingTransfomation}
\end{equation}
resulting in the transformed matrix $\mathbf{\hat{\bm{\Psi}}}_{m} \in \mathbb{R}^{N_m^p \times \mathsf{N}_e}$. 

The SPH simulation of the flow case $m$ runs with fixed $\bm{\psi}_m$ and $\bm{\eta}_m$, and generates a sequence of particle positions from $t=0$ to the final discretized time step $t=t_n$, 
which is represented as 
\begin{gather}
\mathbf{X}_m = \left[ \mathbf{X}_m^{t=0}, \mathbf{X}_m^{t=1}, \dots, \mathbf{X}_m^{t=t_n} \right]\\
\text{with} \quad \mathbf{X}_m^{t_k} = \left[ \mathbf{x}_{m(1)}^{t_k}, \mathbf{x}_{m(2)}^{t_k}, \dots, \mathbf{x}_{m(N_p^m)}^{t_k} \right]
\end{gather}
where $\mathbf{x}_{m(i)}^{t_k}$ denotes the position of the $i$-th particle at the time step $t_k$. 
To account for measurement and modeling uncertainties, random noise is added to the velocity domain and then cumulatively summed to produce perturbations in particle position data.
Specifically, velocity noise follows a Gaussian distribution and scales with $\sigma / \sqrt{t_n}$, where $ \sigma $ is the noise standard deviation and $ t_n $ is the total number of time steps:
\begin{equation} 
\mathbf{v}_{\text{noise}}^{t_k} \sim \mathcal{N} \left( 0, \sigma^2/{t_n} \right) 
\end{equation}
These noisy velocity increments are then integrated over time to produce the final position noise:
\begin{equation}
\mathbf{x}_{\text{noise}}^{t_k} = \sum_{j=1}^{k} \mathbf{v}_{\text{noise}}^{t_j} 
\end{equation}

To enable a GNN model that learns the temporal evolution of particle motion, we consider the velocity subsequences as input features, which are prepared as follows. 
First, the SPH-generated position data $\mathbf{X}_{m}$ are segmented into the position subsequences $\mathbf{X}_m^{t_{k-w}:t_{k-1}}$ from time step $t_{k-w}$ to $t_{k-1}$ by moving a sliding window of size $w$ along the temporal axis of $\mathbf{X}_m$. Then, the corresponding velocity subsequences $\mathbf{s}_{m, t_k}^v \in \mathbb{R}^{N_p^m \times (w-1)d}$ are computed from difference between consecutive positions: 
\begin{equation}
\mathbf{s}_{m, t_k}^v=\mathbf{X}_m^{t_{k-w+1}:t_{k-1}} - \mathbf{X}_m^{t_{k-w}:t_{k-2}}.
\end{equation}

The GNN model is tasked with predicting particle acceleration—and consequently velocity—at the target time $t_k$, using the velocities from the preceding $w$ time steps as temporal context.

\begin{figure}
\begin{center}
\includegraphics[width=0.65\textwidth]{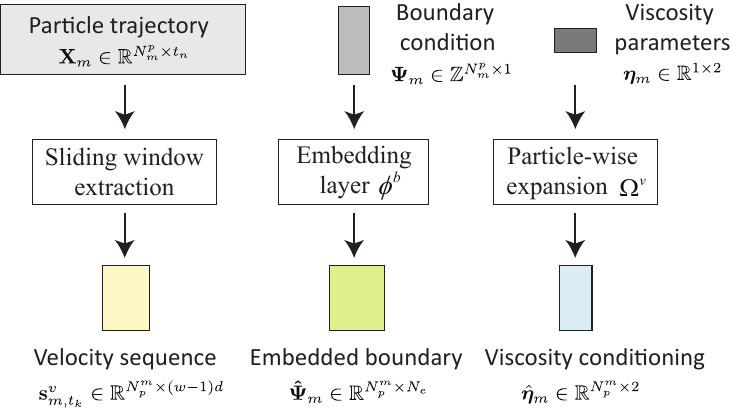}
\end{center}
\caption{Schematic diagram of feature embedding process.
}
\label{fig:Model_feature}
\end{figure}

Finally, the viscosity parameter $\bm{\eta}_{m}$ is expanded into a particle-wise viscosity conditioning matrix through a deterministic transformation $\Omega^v$:
\begin{equation}
\Omega^v: \bm{\eta}_m \to \hat{\bm{\eta}}_{m} \in \mathbb{R}^{N_m^p \times 2}
\label{eqn:viscosityfunction}
\end{equation}
For fluid particles, viscosity values were normalized through min-max scaling based on global minima and maxima across all training simulation cases. For rigid particles, a constant value of 1 was assigned to clearly differentiate them from the normalized fluid particles.
Unlike previous studies~\cite{Sanchez-Gonzalez2020,liGraphNeuralNetworkaccelerated2022a,zhangHybridMethodCombining2024} that rely on categorical particle labels representing pre-defined fluid types, our approach directly incorporates continuous viscosity parameters into the feature embedding process. These embedded features subsequently serve as the node attributes used for constructing the graph representation. Consequently, our model can effectively generalize fluid behavior prediction across diverse non-Newtonian fluid conditions without requiring discrete type labels or re-training.
The overall feature embedding procedure is summarized in \fref{fig:Model_feature}.

\subsection{Graph Construction}

\begin{figure}
\begin{center}
\includegraphics[width=0.65\textwidth]{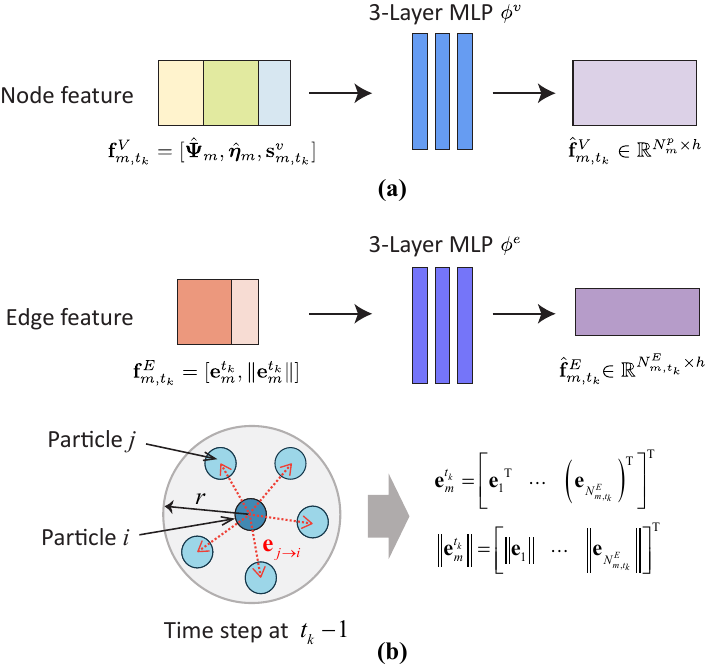}
\end{center}
\caption{Graph construction process: (a) node feature and (b) edge feature structures.}
\label{fig:Model_feature2}
\end{figure}

In this section, we present the GNN architecture used in this study.
Our model architecture builds on the message-passing mechanisms previously implemented in graph network (GN) frameworks \cite{battaglia2018relational, sanchez2018graph,Sanchez-Gonzalez2020}.
For each position subsequence, a graph structure is constructed based on the particle state of $\mathbf{X}_m^{t_{k-1}}$, the immediate previous step of the target time step $t_k$. The edges are established between two particles that lie within a predefined connectivity radius $r$. For each edge connecting a receiving node $i$ and one of its neighboring nodes $j$, an edge vector $\mathbf{e}_{j \rightarrow i}$ is computed as $\mathbf{e}_{j \rightarrow i}=\mathbf{x}_{i}-\mathbf{x}_{j}$.
These edge vectors are structured into the matrix form $\mathbf{e}_m^{t_k} \in \mathbb{R}^{N_{m, t_k}^E \times d}$ along with their magnitude $\| \mathbf{e}_m^{t_k} \|$:
\begin{equation*}
\mathbf{e}_m^{t_k} = \left[ \mathbf{e}_1^\top, \dots, \left( \mathbf{e}_{N_{m,t_k}^E} \right)^\top \right]^\top,
\end{equation*}
\begin{equation*}
\| \mathbf{e}_m^{t_k} \| = \left[ \| \mathbf{e}_1 \|, \dots, \|\mathbf{e}_{N_{m,t_k}^E} \| \right]^\top,
\end{equation*}
where $N_{m, t_k}^E$ is the number of edges and $d$ denotes the spatial dimension.

Next, the node and edge feature vectors are constructed to characterize particle-wise properties (node feature $\mathbf{f}_{m, t_k}^V$) and their pairwise interactions (edge feature $\mathbf{f}_{m, t_k}^E$) respectively: 
\begin{equation*}
\mathbf{f}_{m, t_k}^V = [\hat{\bm{\Psi}}_{m}, \hat{\bm{\eta}}_{m}, \mathbf{s}_{m, t_k}^v],
\end{equation*}
\begin{equation*}
\mathbf{f}_{m, t_k}^E = [\mathbf{e}_m^{t_k}, \| \mathbf{e}_m^{t_k} \|],
\end{equation*}
where $\mathbf{f}^V \in \mathbb{R}^{N^p_m \times h_v}$ with $h_v=N_e+(w-1)d+2$ and $\mathbf{f}^E \in \mathbb{R}^{N^E_m \times h_e}$ with $h_e=d+1$. These features are then encoded into latent representations through learnable neural networks $\phi^v$ and $\phi^e$:
\begin{equation}
{\phi}^v: \mathbb{R}^{h_v} \rightarrow \mathbb{R}^h, \quad \hat{\mathbf{f}}_{m, t_k}^V={\phi}^v(\mathbf{f}_{m, t_k}^V) \in \mathbb{R}^{N_m^p \times h}
\label{eqn:latent_space}
\end{equation}
\begin{equation}
{\phi}^e: \mathbb{R}^{h_e} \rightarrow \mathbb{R}^h, \quad \hat{\mathbf{f}}_{m, t_k}^E={\phi}^e(\mathbf{f}_{m, t_k}^E) \in \mathbb{R}^{N_{m, t_k}^E \times h},
\end{equation}
in which $h$ denotes the latent feature dimension (hidden dimension).
Then, the latent graph representation ${G}$ is structured as
\begin{equation}
G=(V, E), \quad \rm{with} \hat{\mathbf{f}}_{m, t_k}^V \in V, \; \rm{and} \; \hat{\mathbf{f}}_{m, t_k}^E \in E, 
\end{equation}
where each element in vertex set V and edge set E are characterized by their respective learned embeddings. 

The graph construction procedure is summarized in \fref{fig:Model_feature2}. 
In this study, we adopted 3-layer multilayer perceptrons (MLPs) for both $\phi^v$ and $\phi^e$.
Each MLP employs ReLU activation functions after every hidden layer and  the layer normalization is applied to all layers to stabilize training, while a consistent hidden layer dimension across all layers is maintained.

\subsection{Interaction via Message Passing}
\label{subsec:message}

\begin{figure}
\begin{center}
\includegraphics[width=0.65\textwidth]{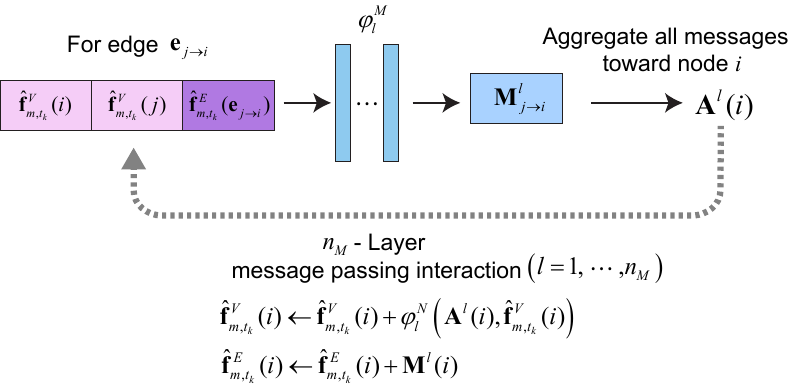}
\end{center}
\caption{Overview of the message passing process at the $l$-th interaction layer.}
\label{fig:Model_MP}
\end{figure}

In the interaction network, the latent graph representations are iteratively refined through structured message passing operations. 
The message $\mathbf{M}^l_{j \rightarrow i}$ is created between each $ij$-pair of nodes connected by $\mathbf{e}_{j \rightarrow i}$
\begin{equation}
\mathbf{M}^l_{j \rightarrow i} = {\varphi}^M_l\left(\hat{\mathbf{f}}_{m, t_k}^V(i), \hat{\mathbf{f}}_{m, t_k}^V(j), \hat{\mathbf{f}}_{m, t_k}^E(\mathbf{e}_{j \rightarrow i})\right) 
\label{eqn:message}
\end{equation}
where ${\varphi}^M_l$ is a $n_\varphi$-layer MLP that learns to generate messages between nodes. The resulting output message is $\mathbf{M}^l_{j \rightarrow i} \in \mathbb{R}^{N_{m, t_k}^E \times h}$ at the $l$-th interaction network layer. 
By summing all incoming messages directed to the same receiving node $i$, the generated messages are aggregated into the aggregation matrix $\mathbf{A}^l$, 
\begin{equation}
\mathbf{A}^l(i) =\sum_{j \in \mathcal{N}(i)} \mathbf{M}^l_{j \rightarrow i} \in \mathbb{R}^{N_{m}^p \times h}
\end{equation}
where $\mathcal{N}(i)$ represents the set of neighboring nodes connected to $i$-th node within a specified connectivity radius $r$. 
The node features are updated by processing both the aggregation matrix $\mathbf{A}^l$ and the current node features $\hat{\mathbf{f}}_{m, t_k}^V$ through a residual connection: 
\begin{equation}
\hat{\mathbf{f}}_{m, t_k}^V(i) \leftarrow \hat{\mathbf{f}}_{m, t_k}^{V}(i) + {\varphi}^N_l(\mathbf{A}^l(i), \hat{\mathbf{f}}_{m, t_k}^V(i))
\end{equation}
where ${\varphi}^N_l$ is a learned function of $n_\varphi$-layer of MLP ${\varphi}^N_l: \mathbb{R}^{h} \rightarrow \mathbb{R}^h$.
The edge features are updated as
\begin{equation}
\hat{\mathbf{f}}_{m, t_k}^E(i) \leftarrow \hat{\mathbf{f}}_{m, t_k}^E(i) + \mathbf{M}^l(i).
\end{equation}
This process iterates for a $n_{M}$ number of layers, updating the latent graph representations.
The final node embeddings are then processed through an output function ${\phi}^O$, which generates the predicted acceleration $\mathbf{a}_{m, t_k}$
\begin{equation}
{\phi}^O: \mathbb{R}^{h} \rightarrow \mathbb{R}^d, \quad  \mathbf{a}_{m, t_k}={\phi}^O(\hat{\mathbf{f}}_{m, t_k}^V) \in \mathbb{R}^{N_{m}^p \times d}.
\end{equation}
The overall process of message generation, aggregation, and iterative feature update is illustrated in \fref{fig:Model_MP}.

\subsection{Learning Process}

Given the available training dataset $\{\mathbf{X}_{m},\bm{\Psi}_{m}, \bm{\eta}_{m}\}_{m=1}^{N}$,
the GNN model learns a simulator function $ \boldsymbol{S}_{\theta}$
with the learnable parameters of the model $\theta$, 
to predict the full sequence of particle accelerations $\tilde{\mathbf{a}}_{m}$ from the given initial particle positions $\mathbf{X}_{m}^0$, the boundary conditions \( \bm{\Psi}_{m} \) and the viscosity parameters \( \bm{\eta}_{m} \), such that:
\begin{equation}
\boldsymbol{S}_{\theta}: (\mathbf{X}_{m}^0, \bm{\Psi}_{m}, \bm{\eta}_{m}) \mapsto \tilde{\mathbf{a}}_{m}.
\end{equation}
Note that we use the tilde symbol ($\tilde{}$) to denote the quantities predicted by the GNN.
This learning process is formulated as an optimization problem, where $\theta$ is learned by minimizing the loss function \( \mathcal{L}(\theta ; \mathbf{a}_{m}, \tilde{\mathbf{a}}_{m}, \mathbf{X}_{m}, \tilde{\mathbf{X}}_{m}) \):
\begin{equation} \label{learning}
    \boldsymbol{S}_{\theta} = \underset{\boldsymbol{S}_{\theta}}{\mbox{argmin}} \sum_{m=1}^{N} \mathcal{L}( \theta; \mathbf{a}_{m}, \tilde{\mathbf{a}}_{m}, \mathbf{X}_{m}, \tilde{\mathbf{X}}_{m}),
\end{equation}

where $\tilde{\mathbf{a}}_{m}$ and $\tilde{\mathbf{X}}_{m}$ denote the predicted accelerations and predicted position sequences, respectively.
To train the GNN-based power-law fluid simulation model, we employ a multi-term loss function that balances acceleration accuracy, long-term position stability, and macroscopic spatial coherence. The loss function \( \mathcal{L}(\boldsymbol{S}_{\theta}) \) consists of three primary terms:
\begin{equation} \label{loss}
    \mathcal{L}(\mathbf{a}_{m}, \tilde{\mathbf{a}}_{m}, \mathbf{X}_{m}, \tilde{\mathbf{X}}_{m}) = \mathcal{L}_{\mathrm{MSE}} +\mathcal{L}_{\mathrm{ACD}},
\end{equation}
with
\begin{equation}
\mathcal{L}_{\mathrm{MSE}} = \frac{1}{N} \frac{1}{N_m^p} \sum_{m=1}^{N} \sum_{i=1}^{N_m^p} \|\mathbf{a}_{m,i} - \tilde{\mathbf{a}}_{m,i}, \|_2^2,
\end{equation}
\begin{equation}
\mathcal{L}_{\textit{ACD}} = \frac{1}{N} \sum_{m=1}^{N}\frac{1}{|\mathbf{X}_{m}|} \sum_{\mathbf{x} \in \mathbf{X}_{m}} \min_{\tilde{\mathbf{x}} \in \tilde{\mathbf{X}}_{m}} \|\mathbf{x}-\tilde{\mathbf{x}}\|_2.
\end{equation}

First, the acceleration loss $\mathcal{L}_{\mathrm{MSE}}$ ensures accurate prediction of instantaneous particle dynamics by minimizing the mean squared error (MSE) between predicted $\tilde{\mathbf{a}}_{i}$ and ground truth accelerations $\mathbf{a}_{i}$. Since the acceleration is the direct output of the learned model, this term is understood to guide the model to accurately capture the underlying physical forces governing particle movement.
On the other hand, the asymmetric Chamfer Distance loss $\mathcal{L}_{\mathrm{ACD}}$~\cite{fan2017point} evaluates the overall similarity between predicted and ground truth particle distributions. Unlike MSE-based losses that enforce strict particle-to-particle correspondence, this term helps maintain macroscopic spatial coherence.  

When the GNN model predicts the acceleration $\tilde{\mathbf{a}}$ starting from an initial state, the predicted acceleration is then used to update the velocity and position, and the updated position serves as input for predicting the next acceleration.  
This process is recursively repeated until the final time step $t_n$, using $\tilde{\mathbf{v}}_{t+1} = \tilde{\mathbf{v}}_t + \tilde{\mathbf{a}}_t$, $\tilde{\mathbf{x}}_{t+1} = \tilde{\mathbf{x}}_t + \tilde{\mathbf{v}}_{t+1}$.
The position loss is computed by comparing the full predicted trajectory $\tilde{\mathbf{X}}_{m}$ with the ground truth trajectory $\mathbf{X}_{m}$.

We determined the initial hyperparameters based on the literature \cite{sanchez2018graph,Sanchez-Gonzalez2020}, including sliding window size $w=7$, connectivity radius $r=0.015$ (which is equal to the kernel function radius used in SPH), embedding dimension $\mathsf{N}_e=16$, the number of MLP layers $n_\varphi=3$, number of message passing layers $n_{M}=10$, and latent feature dimension $h=128$. The network was implemented using Python with PyTorch, where MLPs were constructed using PyTorch's built-in modules, and the message passing network was implemented using PyTorch Geometric. The weights were initialized using the Xavier initializer~\cite{pmlr-v9-glorot10a}. The model was trained using the Adam optimizer for 100 epochs \cite{kingma2017adammethodstochasticoptimization}, with a batch size of 4 and an initial learning rate of 0.0001. An ExponentialLR scheduler~\cite{scheduler} was employed to gradually decay the learning rate over training steps. The implementation was based on PyTorch 2.5.1 with CUDA 12.4. 

\subsection{Data Generation}

\begin{figure}
    \begin{center}
        \includegraphics[width=0.9\textwidth]{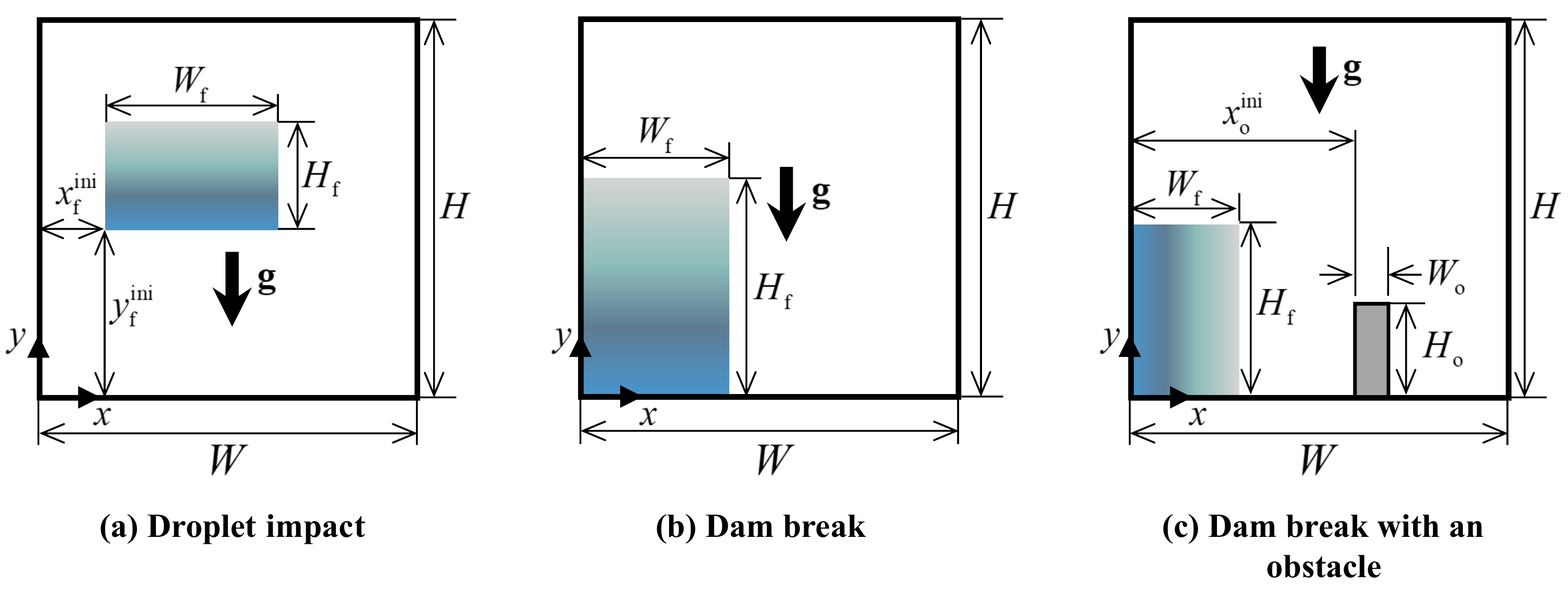}
        \caption{Simulation domains used for benchmark tests: (a) Droplet impact problem, (b) Dam break problem, and (c) Dam break with an obstacle problem.}
        \label{fig:problem_domain_all}
    \end{center}
\end{figure}

\begin{figure}
\begin{center}
\includegraphics[width=0.7\textwidth]{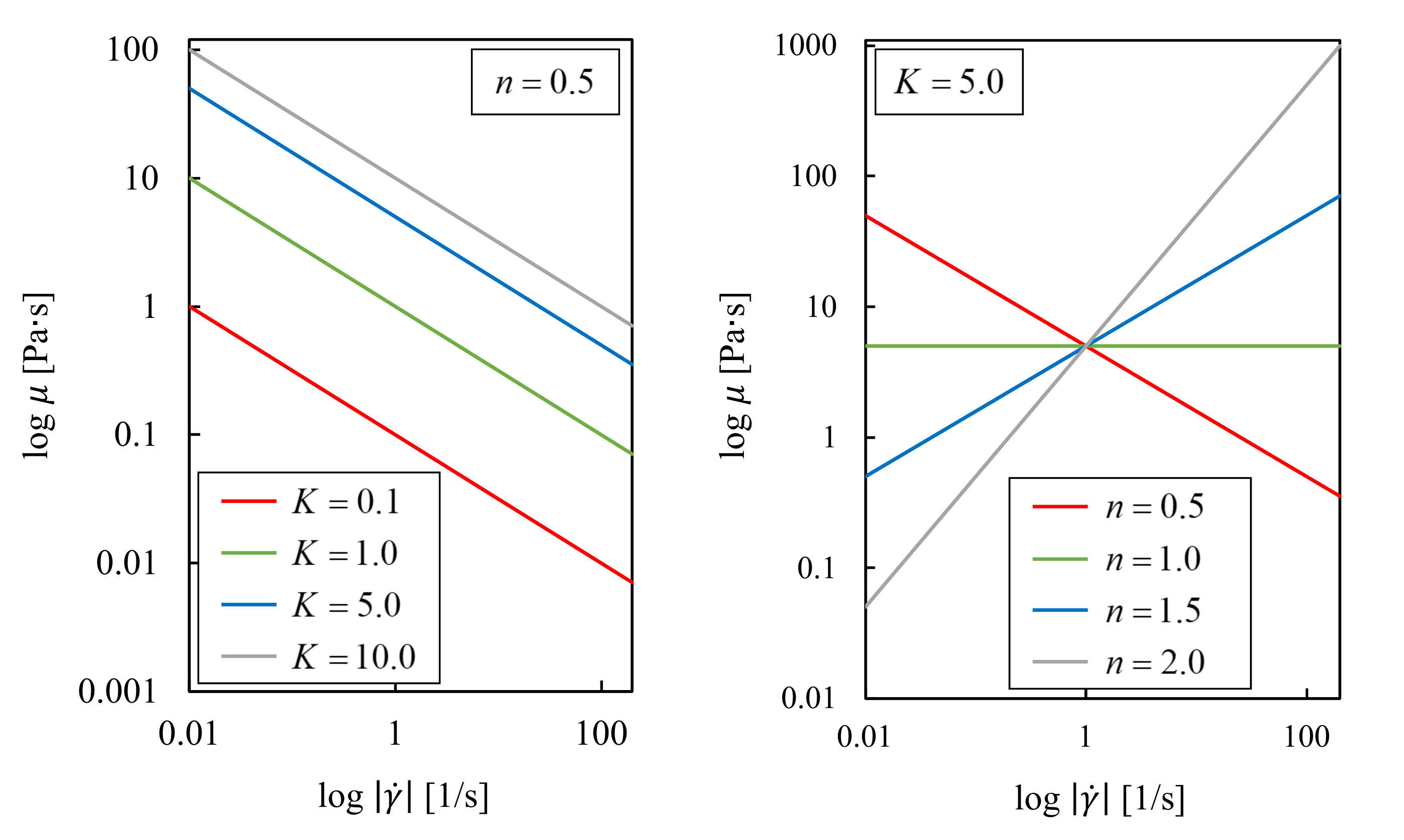}
\end{center}
\caption{Viscosity–shear rate relationships in power-law fluids under varying model parameters. (Left) Viscosity profiles for different flow consistency indices $K$ = 0.1, 1.0, 5.0, 10.0 Pa$\cdot$s at a fixed flow behavior index $n$ = 0.5. (Right) Viscosity profiles for different flow behavior indices $n$ = 0.5, 1.0, 1.5, 2.0 at a fixed consistency index $K$ = 5.0 Pa$\cdot$s. Both plots use logarithmic scales.
}\label{fig:flow_consistency_behavior_index}
\end{figure}

For the training dataset, we create 100 scenes by randomly placing fluid blocks, rectangular rigid body obstacles in a cubic box domain. The dataset is generated from three scenarios: droplet impact, dam break, and dam break with an obstacle. The descriptions of each scenario are as follows:

(1) Droplet impact: as illustrated in \fref{fig:problem_domain_all}(a), a fluid box with an initial width $W_\text{f}$ and height $H_\text{f}$ is placed at a random location ($x^{\text{ini}}_{\text{f}}$,$y^{\text{ini}}_{\text{f}}$) within a rectangular domain of horizontal length $W=0.8\,\rm{m}$ and vertical length $H = 0.8\,\rm{m}$. The size of the fluid box, $W_\text{f}$ and $H_\text{f}$, is randomly determined within the range of 0.15 to 0.3 m, and its initial position is randomly placed at a minimum height of 0.3 m or higher. This scenario is designed to simulate fluid behavior upon impact with a solid surface.

(2) Dam break: this scenario involves a fluid tank of horizontal length $W=0.8\,\rm{m}$ and vertical length $H = 0.8\,\rm{m}$ containing a rectangular fluid column. The dimensions of the fluid column are defined by its width $W_{\rm f}$ and height $H_{\rm f}$, as illustrated in \fref{fig:problem_domain_all}(b). Here, $W_{\rm f}$ and $H_{\rm f}$ are randomly determined within the range of 0.2 to 0.4 m. In this scenario, the initial fluid column collapses under gravity as the simulation progresses, naturally modeling the free surface flow based on the positions of the outermost particles.

(3) Dambreak with an obstacle: the scenario domain consists of a fluid column with a width of $W_{\rm f}$ and a height of $H_{\rm f}$, along with an obstacle positioned at $x_{\text{o}}^{\text{ini}}$ from the origin, with a width of $W_{\rm o}$ and a height of $H_{\rm o}$, as shown in \fref{fig:problem_domain_all}(c). $W_{\rm f}$ and $H_{\rm f}$ are randomly determined within the range of 0.2 to 0.4 m. The position of the obstacle, $x_{\text{o}}^{\text{ini}}$, is randomly set within a region at least 0.1 m away from the end of the fluid column and at least 0.1 m from the right wall. Additionally, the width of the obstacle, $W_{\rm o}$, is randomly determined within the range of 0.01 to 0.05 m, and its height, $H_{\rm o}$, is set randomly between 0.1 and 0.2 m. The fluid column collapses under gravity and impacts the obstacle, exhibiting flow patterns that depend on the fluid's rheological properties. This scenario is designed to validate the interaction between fluid particles and the obstacle, which is modeled using boundary particles.

The gravitational acceleration is set to 9.81 $\rm m/s^2$, and the fluid density is 1000 $\rm kg/m^3$ for the entire scenario. In this paper, the range of $K$ is limited to 0.1–10.0 Pa$\cdot$s, and the range of $n$ is limited to 0.5–2.0. \fref{fig:flow_consistency_behavior_index} illustrates the relationship between shear rate and viscosity for different values of $K$ and $n$.
These restrictions not only exclude fluids that are difficult to represent using the power-law model, but also reduce the need for an excessively large dataset.
The generated data were partitioned into training (80 scenes), validation (10 scenes), and test (10 scenes) sets.

\section{Results and Discussion}
\label{sec:result}

\subsection{Generalization Performance on Randomized Test Cases}
\label{sec:result(a)}

This section evaluates the generalization performance of the GNN-based model using 10 randomly sampled test cases across three representative benchmark scenarios: droplet impact, dam break, and dam break with an obstacle. Each test case varies in initial fluid height, volume, and viscosity parameters, reflecting diverse physical conditions.
The evaluation is presented in two parts: (i) stepwise MSE trends to assess temporal prediction accuracy and (ii) correlation between cumulative MSE and key parameters to explore sensitivity.

\subsubsection{Stepwise Error Trends}
\label{sec:Stepwise_MSE}

\begin{figure}
\begin{center}
\includegraphics[width=0.6\textwidth]{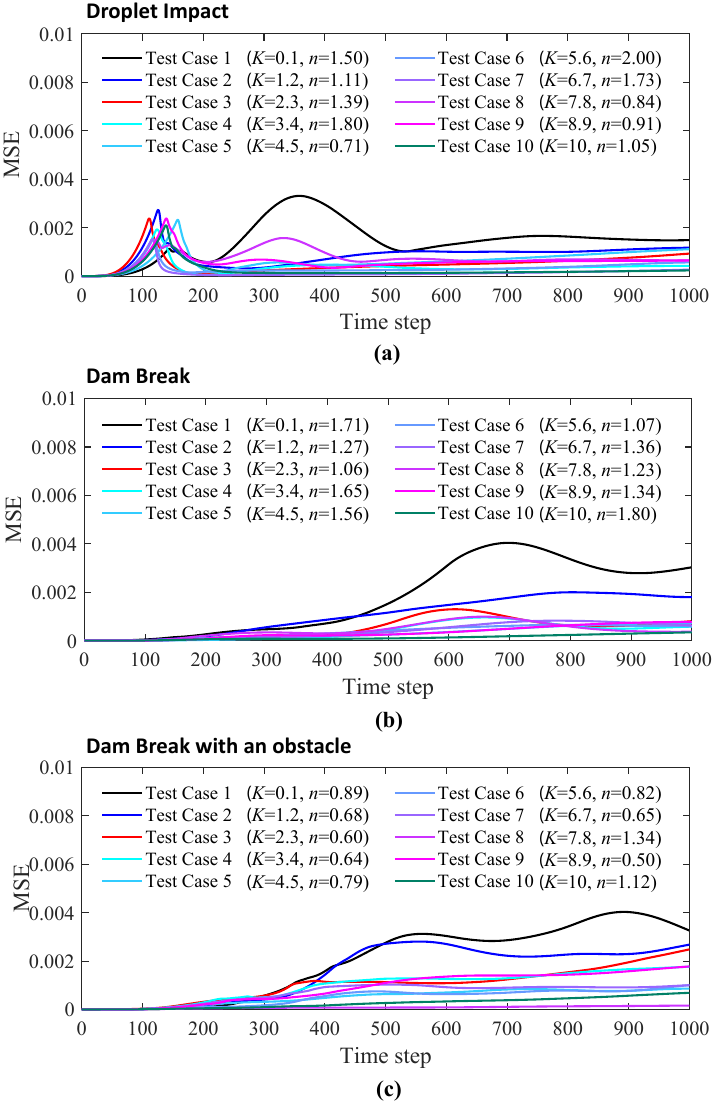}
\end{center}
\caption{Stepwise MSE over total time steps: (a) droplet impact, (b) dam break, and (c) dam break with an obstacle.}
\label{fig:Stepwise_MSE}
\end{figure}

\begin{figure}
\begin{center}
\includegraphics[width=0.8\textwidth]{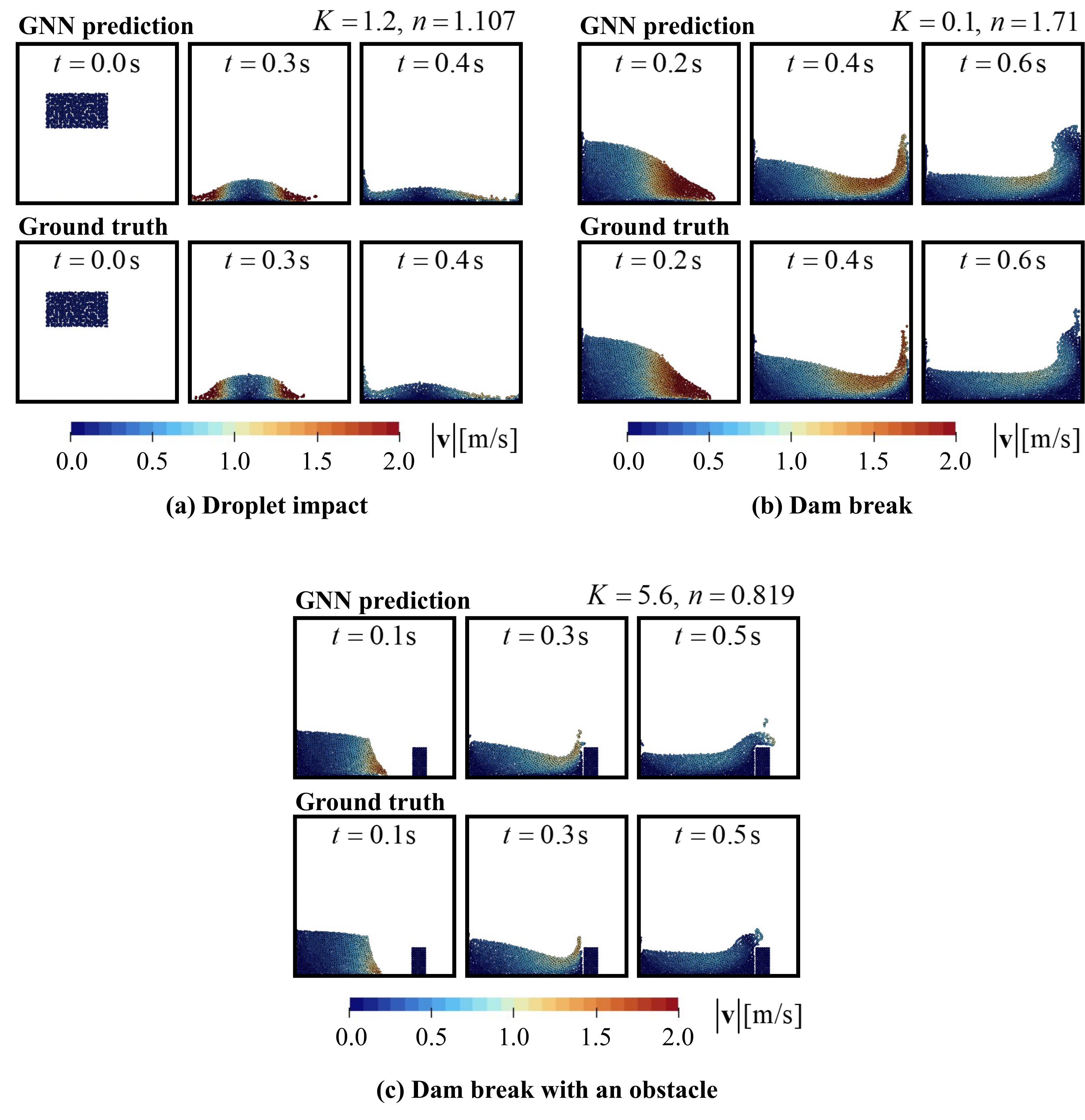}
\end{center}
\caption{Simulation snapshot comparisons between the GNN prediction and the ground truth: (a) droplet impact ($W_f=0.294$ m, $H_f=0.164$ m, $x^{\text{ini}}_{\text{f}}=0.157$ m, $y^{\text{ini}}_{\text{f}} = 0.377$ m), (b) dam break ( $W_{\text{f}}=0.344$ m, $H_{\text{f}}=0.304$), and (c) dam break with an obstacle ($W_{\rm f}=0.343$ m, $H_{\rm f}=0.234$ m, $W_{\rm o}=0.06$ m, $H_{\rm o}=0.14$ m, and $x_{\text{o}}^{\text{ini}}=0.686$ m). }
\label{fig:snapshots_allcases}
\end{figure}

\fref{fig:Stepwise_MSE} shows the time step-wise MSE across 10 independently sampled test cases for the three benchmark scenarios. The test cases are ordered by increasing flow consistency index ($K$), enabling a clear examination of how viscosity affects prediction accuracy. A consistent trend emerges across all scenarios: lower-$K$ cases result in significantly higher prediction errors. This suggests the GNN model has difficulty capturing rapidly deforming, less viscous flows.

In the droplet impact scenario (\fref{fig:Stepwise_MSE}(a)), most test cases exhibit a sharp MSE peak during the early time steps (100–200), coinciding with the initial droplet–surface interaction. This peak is followed by a decline, indicating stabilization of the fluid configuration.
In contrast, the dam break scenario (\fref{fig:Stepwise_MSE}(b)) shows a gradual increase in MSE over time, particularly in low-$K$ cases, pointing to error accumulation as the fluid spreads and interacts with boundaries. 
The dam break with obstacle scenario (\fref{fig:Stepwise_MSE}(c)) displays a similar upward trend in prediction error, particularly in low-$K$ cases, where the model appears more sensitive to chaotic flow structures and fluid–obstacle interactions.
Overall, the results demonstrate a strong sensitivity of prediction accuracy to viscosity-related parameters, particularly in scenarios involving rapid flow transitions or complex interactions with boundaries.

\fref{fig:snapshots_allcases} presents qualitative comparisons between the predicted and ground-truth particle distributions for representative test cases in each scenario. Each snapshot shows particle positions at three time steps with particle color indicating velocity magnitude. Overall, the GNN predictions exhibit flow patterns that are qualitatively consistent with the ground truth, capturing key features of the fluid configuration and motion. Localized differences such as earlier spreading or smoothed velocity gradients can be observed in some cases, particularly near obstacles or rapidly evolving region.

\subsubsection{Influence of Initial Conditions on Prediction Error}

\begin{figure}[h]
\begin{center}
\includegraphics[width=0.6\textwidth]{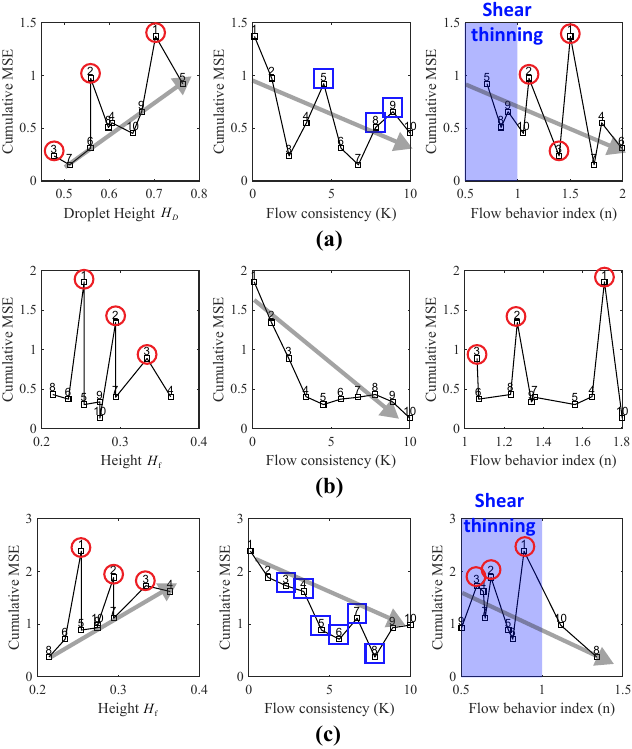}
\end{center}
\caption{Cumulative MSE plots for (a) droplet impact, (b) dam break, and (c) dam break with an obstacle case, respectively where $H_D=y_o^{\text{ini}}+0.5H_f$ for the droplet impact 
and $H_D=H_f$ for other two scenarios. Cases marked with red circles correspond to low-$K$ outliers excluded from the analysis, and blue squares indicate shear-thinning cases ($n<1$).}
\label{fig:Cumulative_MSE}
\end{figure}

To quantify the influence of the initial conditions on prediction accuracy, we computed the cumulative MSE for 30 randomly generated test cases (10 for each benchmark scenario) in Section~\ref{sec:Stepwise_MSE}.
The resulting errors were then plotted against three key parameters: fluid drop height $H$, flow-consistency index $K$, and flow-behavior index $n$ as shown in \fref{fig:Cumulative_MSE}. Note that we excluded (i) the three low-$K$ outliers and (ii) shear-thinning cases ($n<1$) whenever they obscured the individual contribution of $K$, in order to isolate the effect of each parameter.

The cumulative MSE was analyzed excluding the three low-$K$ cases. For the droplet impact and dam break with an obstacle scenarios, cumulative MSE showed a positive correlation with the fluid drop height $H_D$, indicating higher prediction errors for greater drop heights. This can be attributed to increased deformation in the fluid flow due to stronger impacts and rapid variations in shear rates, leading to more complex viscosity-driven behaviors and, consequently, higher prediction errors. Conversely, the dam break scenario exhibited minimal variation in cumulative MSE with respect to fluid height, likely due to the comparatively weaker impact and shear rate variations inherent in this scenario.

For flow-consistency index $K$, shear-thinning cases ($n<1$, blue square markers in \fref{fig:Cumulative_MSE}) were omitted to isolate 
low-$K$ effect. With those cases removed, cumulative MSE shows generally decreased with increasing $K$. 
Higher $K$ values correspond to higher apparent viscosities, reducing the magnitude of flow deformations and thus prediction errors. This inverse relationship was particularly prominent in the dam break scenario, which involves relatively lower impact forces. However, the droplet impact and dam break with obstacle scenarios exhibited some variability in this trend, presumably due to stronger and more chaotic flow behaviors induced by substantial initial impacts and interactions with obstacles.

Excluding the low-$K$ outlier cases, cumulative MSE generally exhibited a mild negative trend with increasing $n$ in both droplet impact and dam break with an obstacle scenarios, indicating slightly reduced prediction errors as $n$ increased. In contrast, the dam break scenario showed negligible sensitivity of cumulative MSE to variations in $n$. This is a natural outcome, considering that in scenarios with relatively large impact forces, an increase in the value of 
$n$ shifts the flow behavior toward shear-thickening.

\subsection{Evaluation under Controlled Viscosity Conditions}
\label{sec:result(b)}

While the previous section focused on evaluating model generalization across randomized test cases, this section aims to more systematically investigate the model’s response to variations in viscosity-related parameters. To isolate the effect of viscosity, additional simulations were conducted under fixed initial conditions, varying only the flow consistency index $
K$ and flow behavior index $n$ across nine combinations. This controlled setup enables a focused assessment of the model’s capacity to capture non-Newtonian characteristics.
Each case was evaluated using the asymmetric Chamfer distance (ACD), computed over time to quantify the spatial discrepancy between predicted and ground-truth particle positions. Additionally, qualitative comparisons of free surface configurations were included to provide visual insight into prediction accuracy.
The results are presented in the following subsections for the three representative scenarios: droplet impact, dam break, and dam break with an obstacle.

\subsubsection{Droplet Impact}

\begin{figure}
    \centering
    \includegraphics[width=0.8\textwidth]{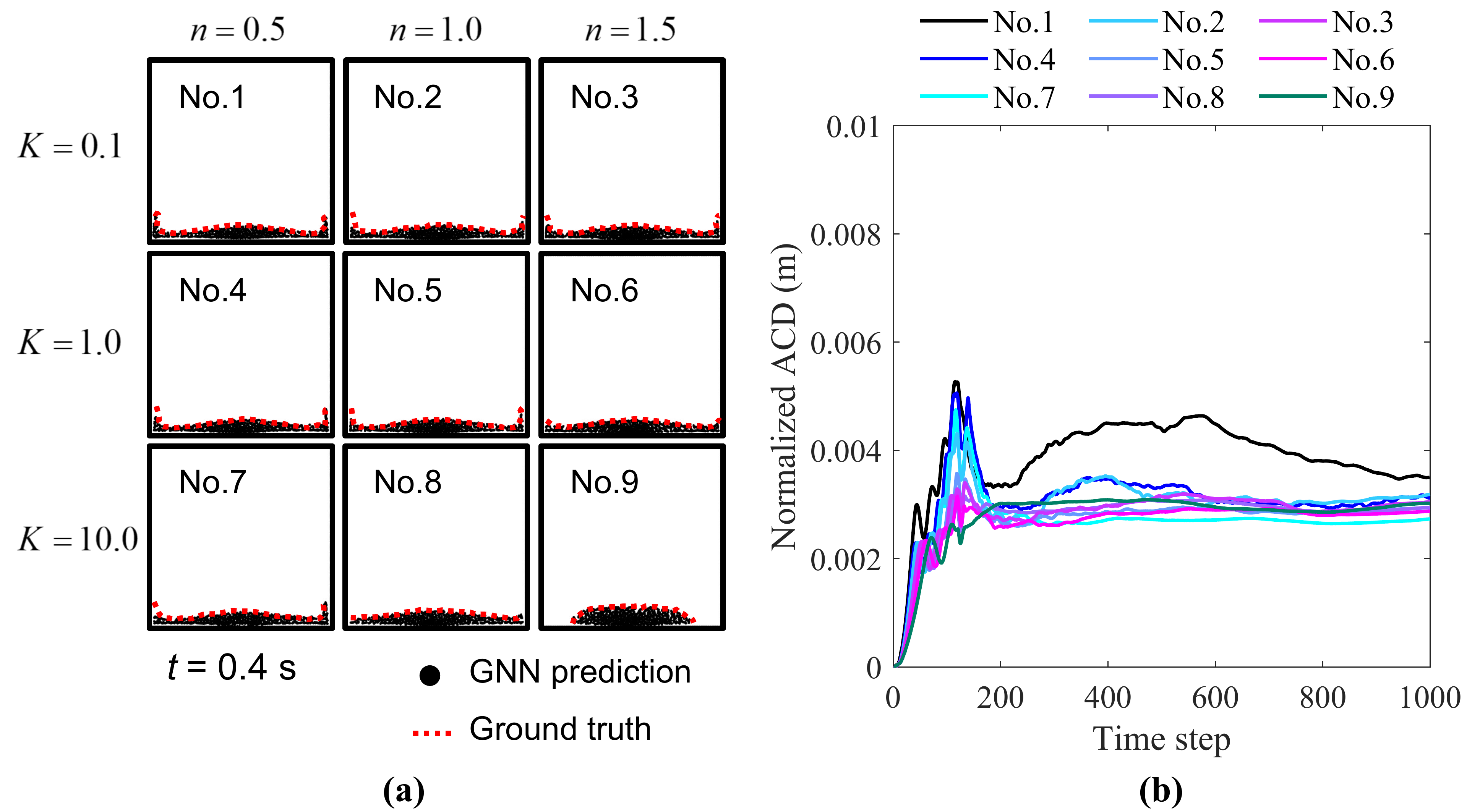}
    \caption{Evaluation of viscosity-dependent prediction accuracy in the droplet impact scenario: (a) comparison of free-surface predictions at $t=0.4$ s across nine viscosity configurations and (b) normalized ACD over time for each viscosity case.}
    \label{fig:Droplet_Impact_9case_viscosity}
\end{figure}

\fref{fig:Droplet_Impact_9case_viscosity} presents the model's performance under nine controlled viscosity settings in the droplet impact scenario. Free-surface predictions at a fixed time step ($t=0.4$ s) are compared in \fref{fig:Droplet_Impact_9case_viscosity}(a), and normalized asymmetric Chamfer distances (ACD) over time are shown in \fref{fig:Droplet_Impact_9case_viscosity}(b). This analysis aims to assess whether the model captures shear-dependent flow behavior, distinguishing between shear-thinning ($n=0.5$) and shear-thickening ($n=1.5$) responses under identical initial conditions.

Across all cases, ACD values remain below 0.005, corresponding to a relative position error of less than 0.5 \% of the characteristic domain length. The ACD plots exhibit a sharp peak near $t=170$, corresponding to the moment of initial surface impact and rapid deformation, followed by a stabilization phase. This transition reflects the fluid's initial contact with the surface, after which the discrepancy plateaus and remains low.

Comparison across viscosity settings reveals that cases exhibiting both shear-thickening behavior and high-viscosity (No.6, No.8, and No.9) result in lower ACD values throughout the simulation, while cases with both shear-thinning and low-viscosity (No.1, No.2, and No.4) tend to exhibit higher discrepancies. Notably, Case No.1 ($K=0.1$, $n=0.5$) shows the highest ACD, suggesting difficulty in modeling rapid spreading dynamics under low-viscosity conditions. These results suggest that the GNN model is more robust in capturing the behavior of high-viscosity fluids, where deformation is more constrained and flow progression is smoother.

\subsubsection{Dam Break}

\begin{figure}
    \centering
    \includegraphics[width=0.8\textwidth]{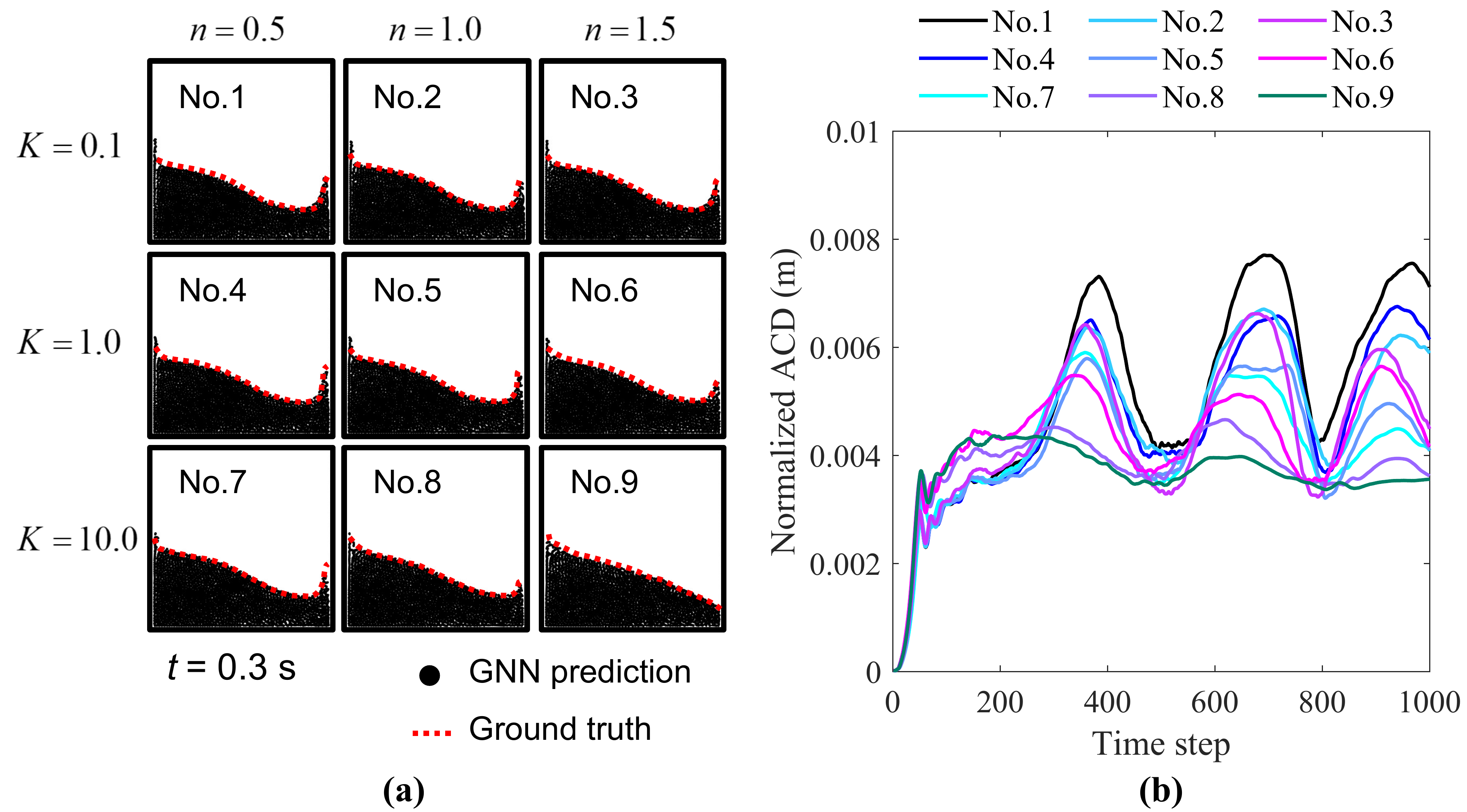}
    \caption{Evaluation of viscosity-dependent prediction accuracy in the dam break scenario: (a) comparison of free-surface predictions at $t=0.3$ s across nine viscosity configurations and (b) normalized ACD over time for each viscosity case.}
    \label{fig:Dambreak_viscosity_9case}
\end{figure}

\fref{fig:Dambreak_viscosity_9case}  shows the model's prediction performance under nine controlled viscosity conditions in the dam break scenario. Free-surface predictions at $t=0.3$ s are visualized in \fref{fig:Dambreak_viscosity_9case}(a), and normalized ACD trends over time are presented in \fref{fig:Dambreak_viscosity_9case}(b). These experiments assess how well the GNN model captures viscosity-dependent flow behavior when the fluid undergoes boundary collisions and large-scale deformation.

The ACD curves show that overall prediction accuracy remains high, with peak discrepancies below 0.008, corresponding to less than 0.8\% of the domain size. However, compared to the droplet impact case, the dam break scenario exhibits greater variability in ACD values over time. In particular, cases with both low-viscosity and shear-thinning behavior—such as No.1, No.2, and No.4—show increased oscillations in ACD, indicating that errors accumulate as the flow repeatedly impacts boundaries and reverses direction. These dynamics amplify initial discrepancies, making prediction more sensitive in such cases.

Conversely, in cases with both high-viscosity and shear-thickening conditions—e.g., No.6, No.8 and No.9—the model predictions remain stable, and the ACD curves show minimal variation after the initial peak. This suggests that stronger resistance to deformation dampens the feedback loop of accumulated prediction errors. Overall, the GNN model maintains high accuracy in predicting viscosity-influenced dam break flows, though performance degrades more rapidly in low-viscosity regimes where error propagation becomes more significant.

\subsubsection{Dam Break with an Obstacle}

\begin{figure}
    \centering
    \includegraphics[width=0.8\textwidth]{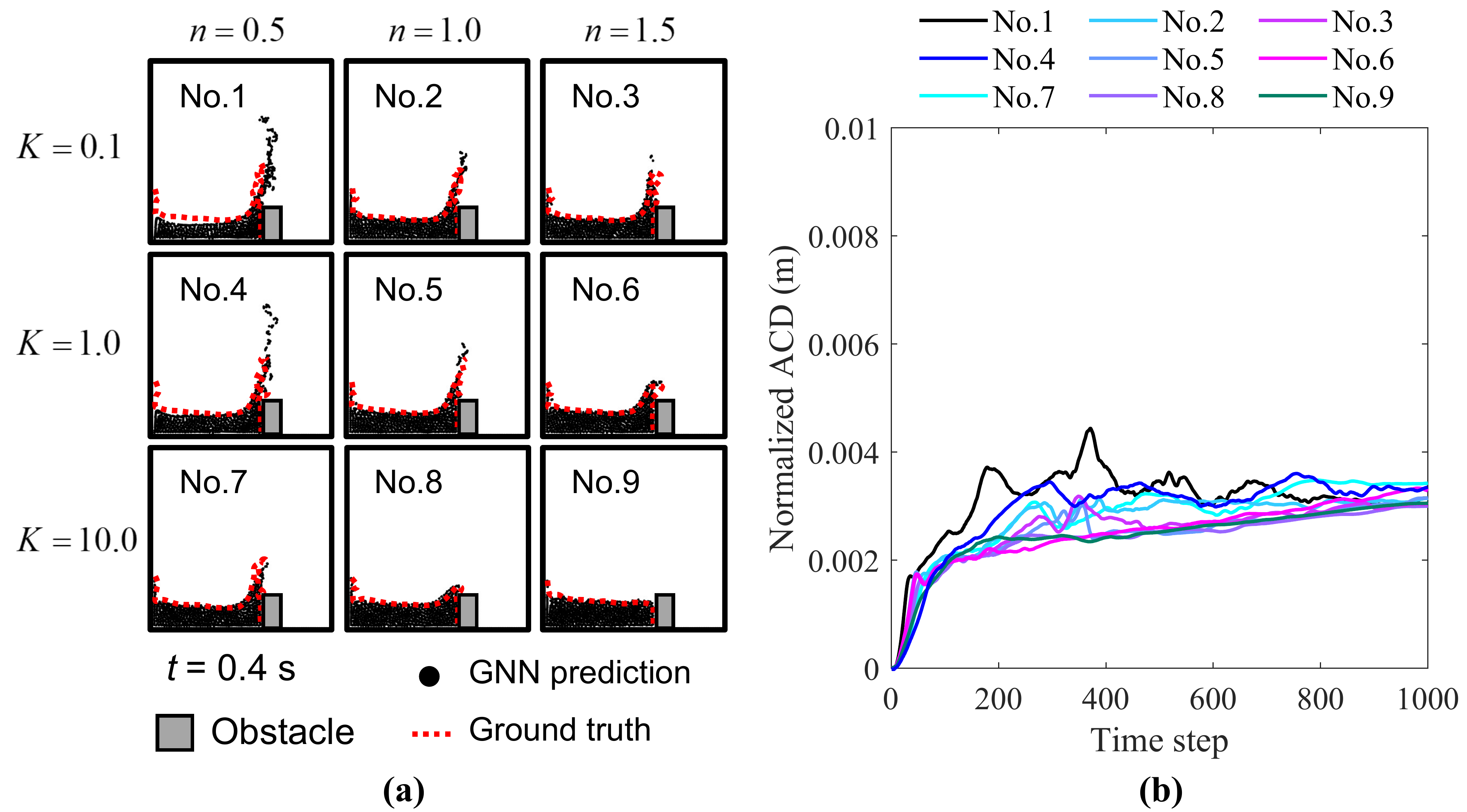}
    \caption{Evaluation of viscosity-dependent prediction accuracy in the dam break with an obstacle scenario: (a) comparison of free-surface predictions at $t=0.4$ s across nine viscosity configurations and (b) normalized ACD over time for each viscosity case.}
    \label{fig:DB_obstacle_viscosity_9case}
\end{figure}

\fref{fig:DB_obstacle_viscosity_9case} shows the model's performance across nine viscosity configurations in the dam break scenario with an obstacle. Particle positions at $t=0.4$ s are shown in \fref{fig:DB_obstacle_viscosity_9case}(a), and corresponding ACD values over time are plotted in \fref{fig:DB_obstacle_viscosity_9case}(b). This setting introduces additional complexity due to obstacle–fluid interactions, providing a more rigorous test of the model's ability to capture viscosity-dependent dynamics.

Compared to the standard dam break case, the presence of the obstacle leads to energy dissipation and flow redirection, resulting in smoother ACD curves with reduced oscillations. Across all cases, ACD remains below 0.005 (less than 0.5\% of the domain size), indicating high predictive accuracy. However, clear differences emerge depending on viscosity. High-viscosity cases with shear-thickening behavior (e.g., No.6–No.9) show lower and more stable ACD values, while low-viscosity cases with shear-thinning behavior (e.g., No.1–No.4) exhibit larger deviations.

In particular, Case No.1 ($K=0.1$, $n=0.5$) again shows the highest ACD values, indicating difficulties in capturing highly dynamic, rapidly spreading flows. In contrast, high-viscosity (shear-thickening) flows remain more cohesive and experience reduced momentum change upon impact, leading to better agreement with the ground truth. These results suggest that the model performs more reliably when flow interactions are stabilized by higher viscosity, effectively capturing large-scale dynamics even under obstacle-induced perturbations.

\subsection{Summary and Discussion}

The results presented in \sref{sec:result(a)} and \sref{sec:result(b)} collectively demonstrate that the proposed GNN-based model achieves high prediction accuracy across a wide range of fluid scenarios, including complex viscosity-dependent behaviors. In general, the model successfully reproduces the overall flow structures and dynamic characteristics observed in the ground truth data, particularly in post-impact regimes.

One notable observation is a slight shift in predicted impact timing compared to the ground truth, which is attributed to fixed gravitational acceleration in the training data. Since the model was not exposed to varying gravitational forces during training, it cannot fully generalize to free-fall conditions. While this limitation leads to minor temporal offsets, it does not significantly affect the model's spatial accuracy and is therefore considered acceptable for the current evaluation.

The influence of viscosity is clearly reflected in the model's performance across all benchmark cases. Shear-thinning fluids ($n<1.0$) exhibit more rapid and unstable flow patterns, resulting in greater spatial discrepancies as measured by the ACD metric. In contrast, shear-thickening fluids ($n>1.0$) tend to suppress flow variations, leading to more cohesive and predictable motion. Similarly, lower $K$ values, which correspond to lower overall viscosity, are associated with higher prediction errors. These observations highlight the model’s increased sensitivity to rapidly evolving, low-viscosity flow regimes where minor discrepancies can accumulate over time.

In terms of computational performance, Table \ref{table_computation_time} shows that the proposed model is significantly more efficient than conventional SPH-based simulations, achieving runtime reductions of approximately 28–33\% across all scenarios. This efficiency gain, combined with strong predictive performance, suggests that the GNN-based approach offers a promising balance between accuracy and computational cost for large-scale, data-driven fluid simulations.

It is important to note that the computational cost for identifying neighboring particles—a necessary step for defining local particle interactions—is equivalent in both the GNN and SPH frameworks. Therefore, the primary difference in computational efficiency arises from the force evaluation stage. In SPH, this involves explicit calculations of pairwise particle forces based on physical models, whereas in the proposed GNN model, these interactions are learned and predicted via message-passing operations. This distinction becomes increasingly important in three-dimensional simulations, where force evaluation accounts for a larger proportion of the total computational cost due to the higher number of particle interactions~\cite{Sanchez-Gonzalez2020}. For this reason, the present study focuses on two-dimensional cases to first validate the efficiency of force interaction modeling. Once validated in 2D, the expected efficiency gains in 3D simulations are likely to be even greater.

\begin{table}[]
\caption{Computational time comparison between GNN and SPH simulations.}
\label{table_computation_time}
\begin{center}
\begin{tabular}{lccc}
\hline
                           & \multicolumn{2}{c}{Computational time [s]} & \multicolumn{1}{l}{}           \\ \cline{2-3} 
Problem domain             & GNN                  & SPH                 & \multicolumn{1}{l}{Efficiency [\%]} \\ \hline
Dam break                  & 462.54   & 671.21   & 31.1                           \\
Droplet impact             & 93.62    & 130.11   & 28.0                           \\
Dam break with an obstacle & 126.75   & 188.58   & 32.8                           \\ \hline
\end{tabular}
\end{center}
\end{table}

\section{Conclusions}
\label{sec:conclusions}

In this study, we developed a GNN model for simulating non-Newtonian free surface flow. Among various models available for non-Newtonian fluid simulation, we adopted the power-law model, which is the most fundamental approach for representing non-Newtonian fluids. Due to its minimal number of parameters, the power-law model allows for efficient learning while effectively determining viscosity.

Since the graph structure of the GNN model is well suited for representing particle interactions in particle-based methods, the training dataset was constructed using Smoothed Particle Hydrodynamics (SPH). To overcome the high computational cost and implementation complexity associated with conventional numerical methods such as SPH, we proposed a GNN-based alternative model.

The training process for the proposed GNN model incorporated fluid particle positions, boundary conditions, and the power-law parameters: the flow consistency index $K$ and the flow behavior index $n$. The trained model successfully captured various flow characteristics depending on the given $K$ and $n$ values, including shear-thinning and shear-thickening effects. To evaluate the model’s performance, we conducted comparisons of free surface evolution, flow trends based on viscosity parameters, and quantitative analysis using the ACD metric. The results demonstrated that the proposed GNN model effectively simulates nonlinear viscosity variations based on the power-law model. Furthermore, compared to the conventional SPH method, the GNN-based model achieved approximately 30.6\% higher computational efficiency, proving that it can serve as a more efficient alternative solution.

This study presents the first GNN-based model for simulating non-Newtonian fluid flow, leveraging an early-stage GNN model originally proposed for particle-based flow simulations~\cite{Sanchez-Gonzalez2020}. Therefore, this research is expected to serve as a foundation for future studies utilizing more advanced GNN models. Additionally, follow-up research will explore more complex non-Newtonian models and various boundary conditions.

\paragraph{Acknowledgments} This research was supported by the Basic Science Research Program through the National Research Foundation of Korea (NRF), funded by the Ministry of Education, grant number 2022R1C1C2006328.
J. Kim also acknowledges support from the National Research Foundation of Korea grant funded by the Korean government under Grant No. RS-2024-00333943.

\paragraph{Author contributions} 
Hyo-Jin Kim: Conceptualization, Software, Methodology, Writing–original draft. 
Jaekwang Kim: Data curation, Writing – review \& editing. 
Hyung-Jun Park: Supervision, Writing–original draft, Writing – review \& editing

\paragraph{Data availability} The data that support the findings of this study are available within the article.

\section*{Declarations}
\paragraph{Competing interests} The authors declare that they have no known competing financial interests or personal relationships that could have appeared to influence the work reported in this paper.

\printbibliography

\end{document}